\documentstyle[twocolumn,aps]{revtex}

\begin{document}
\input{epsf}
\draft
\title{Motional dressed states in a Bose condensate:
Superfluidity at supersonic speed}
\author{C. K. Law, C. M. Chan, P. T. Leung, and M.-C. Chu}
\address{{Department of Physics,
The Chinese University of Hong Kong,}\\
{Shatin, NT, Hong Kong}}
\date{\today}
\maketitle
\begin{abstract}

We present an exact analytic solution of a nonlinear Schr\"odinger
field interacting with a moving potential (obstacle) at supersonic
speed. We show that the field forms a stable shape-invariant
structure localized around the obstacle --- a dressing effect that
protects the field against excitations by the obstacle.

\end{abstract}
%\vspace{10mm}


One of the most intriguing phenomena in superfluid is frictionless
flow below a critical velocity $v_c$ \cite{review}. This
phenomenon, first observed in liquid helium, demonstrates key
features of collective behavior in macroscopic quantum coherent
systems. Recently, experimental evidence of the existence of a
critical velocity in a Bose-Einstein condensed gas has been
found \cite{mit}. It is known that atoms in such a dilute
system interact weakly with each other, and so it justifies
the use of the nonlinear Schr\"odinger equation (NLSE) to study the
quantum dynamics of the condensate. Indeed, direct
numerical solutions of the NLSE have indicated a distinct
transition from superfluid flow to resistive flow at a critical
velocity \cite{rica,adams}. Recent studies have also addressed
dynamic features in solutions of the NLSE, such as phonon and
vortex emissions \cite{nore,jackson,burnett,dalibard}, in order
to understand the dissipation mechanisms.

As an obstacle moves through a Bose condensate, the regime of
supersonic speed is often considered as a dissipation domain where
frictionless motion disappears. This is understood from Landau's
argument that energy-momentum conservation forbids phonon
excitations unless the obstacle's velocity is at least the speed
of sound $c$, i.e., the critical velocity $v_c$ equals $c$ for
systems with a pure phonon spectrum \cite{tim}. In fact, for
realistic systems obeying the NLSE, $v_c$ is {\em less} than $c$
because of the reduced fluid densities appearing at the boundary
of the obstacle \cite{rica,adams,nore,hakim}. Therefore the speed
of sound seems to be an upper limit beyond which dissipation
occurs generally. However, we shall show that there exist
interesting exceptions at least in one dimensional systems. Such
exceptions exhibit rich dynamic features resulted from coherent
field-obstacle interactions, which cannot be captured by the
concept of critical velocities alone.

In this paper we present an exact analytic solution of a nonlinear
Schr\"odinger field interacting with a moving repulsive potential
(obstacle) in one dimension. Our solution represents a family of
{\em motional dressed states} in which the field organizes itself
as a stable and shape invariant structure localized at the moving
obstacle. Once the dressing is fully developed, there is no energy
transfer from the obstacle to the field. We show that this happens
when the velocity of the potential is {\em greater} than the speed
of sound in the field. Therefore dressed states formation is a
novel coherent feature that can be maintained in a condensate at
high speeds. The existence of dressed states at supersonic speed
requires a specific matching between the field and the potential.
Our solution provides a prescription for the matching potentials.
This opens possibilities of coherent control of local properties
of a Bose condensate.

We begin by writing down a nonlinear Schr\"odinger equation for a
field $\Psi (x,t)$ under the influence of a moving potential
$U(x-vt)$, which is well localized and repulsive. As usual,
the NLSE in a dimensionless form is given by
\begin{eqnarray}
i{\partial  \over {\partial t}}\Psi (x,t) &=&
- {1 \over 2}{{\partial ^2 } \over {\partial x^2 }}\Psi (x,t)
+ \left| {\Psi (x,t)} \right|^2 \Psi (x,t) \nonumber \\
&& + U(x - vt)\Psi (x,t)
-\Psi(x,t).
\end{eqnarray}
This equation can be used to determine the evolution of the
condensate wavefunction in one dimension at zero temperature. If
the condensate has a chemical potential $\mu$ and a particle mass
$m$, then the position $x$ and the time $t$ are in units of $\hbar
/ \sqrt {m\mu}$ and $\hbar / \mu$, respectively. In Eq. (1), the
nonlinear term is positive for repulsive self interactions, and
the speed $v$ is taken to be positive for definiteness. We assume
that the field is uniform (i.e., at rest) as $x$ tends to
$\pm\infty$. This imposes the boundary condition
\begin{equation}
\Psi (x \to  \pm \infty ,t) = e^{i\gamma_{\pm}}
\end{equation}
at any finite time $t$. Here $\gamma_{\pm}$ are constants which
account for possibilities of having different phases at the two
infinities. The choice of unit modulus is a convenient scale which
gives the speed of sound $c=1$ for undisturbed systems.

We define a dressed state of the system by a solution of (1) in
the shape invariant form: $\Psi(x,t)=\psi (x-vt)$ subjected to the
boundary condition (2). In other words dressed states are
stationary states in the co-moving frame of the potential. For $v$
in the subsonic domain $v<1$, Hakim has reported stationary states
which are of the form of a dip surrounding the obstacle \cite{hakim}.
The dip appearance is understood because of the repulsive character of
the potential. However, for $v$ greater than the speed of sound,
no known steady states exist. We now describe
a method to determine dressed states in this regime.

By assuming the solution in the form: $\psi(\tau)=
r(\tau)e^{i\theta (\tau)}$, where $\tau \equiv x-vt$, Eq. (1)
requires
\begin{eqnarray}
&& {1 \over 2}\left( {\ddot r - r\dot \theta ^2 } \right) =
\left( {r^3  - r} \right) + U(\tau )r - vr\dot \theta
\\
&&{1 \over 2}\left( {r\ddot \theta  + 2\dot r\dot \theta }
\right)
= v\dot r.
\end{eqnarray}
Here the derivatives are taken with respect to $\tau$, i.e.,
$\dot r = dr/d\tau, \ddot r= d^2 r/d \tau^2$, etc. If we consider
$\tau$ as a kind of time, then
equations (3) and (4) are equivalent to the  motion of a
classical particle interacting with a central force
and a time-dependent magnetic field described by $U(\tau)$.
Therefore a dressed state corresponds to a trajectory
matching the boundary conditions (2). This requires
\begin{eqnarray}
&& r(\tau \to \pm \infty) = 1,  \\
&& \theta (\tau \to \pm \infty) =
\gamma_{\pm}.
\end{eqnarray}
From the constant of motion
$K \equiv vr^2  - r^2 \dot \theta$, we see that $K$
must equal $v$ for desired trajectories. In this way we have
$\dot \theta = v(1-r^{-2})$, and so Eqs. (3) and (4)
are reduced to a one-dimensional trajectory problem \cite{hakim},
\begin{equation}
{1 \over 2}\ddot r =\left( {r^3-r} \right)-{{v^2} \over 2}
\left( {r-{1 \over {r^3}}} \right)+U(\tau )r.
\end{equation}
Once the function $U(\tau)$ is given, we may search for a solution
by finding out initial conditions $r(0)$ and $\dot r(0)$ that make
$r(\tau)$ approach one asymptotically. Although the search can be
performed systematically via numerical means, the existence of a
solution is not a guarantee for arbitrary functions $U$ and speeds
$v$.

We obtain dressed state solutions analytically by adopting an
inverse approach. First we construct an $r(\tau)$ that satisfies
the boundary condition (5), then we determine the corresponding
$U(\tau)$ from Eq. (7). The requirement is that the $U(\tau)$ must
be {\em well localized and everywhere repulsive}, i.e., $U'(\xi
>0) <0$ and $U'(\xi<0) >0$. This approach tells us how one should
design the shape of the $U$ in order to have a specified form
of $\Psi$. The matching of $U(\tau)$ and $r(\tau)$ yields the
exact dressed state solutions. We note that in experimental
situations  \cite{mit}, $U$ can in fact be an adjustable function
controlled by the intensity of a detuned laser.

We give a family of exact solutions in the form of a solitary wave:
\begin{eqnarray}
&& r(\tau )=\left[ {1-{\alpha  \over v} {\rm sech}^2(\beta \tau )}
\right]^{-1/2}
\\
&&
\theta (\tau )={\alpha  \over \beta }\tanh \beta \tau .
\end{eqnarray}
The two {\em positive} parameters $\alpha$ and $\beta$
characterize the
height and inverse width of the field $\Psi$ near the moving
potential. The exact form of $U(\tau)$ that matches Eq. (8) is
given by,
\begin{eqnarray}
U(\tau )={{\alpha {\rm sech}^2(\beta \tau )}
\over {2\left[ {v-\alpha {\rm sech}^2(\beta \tau )}
\right]^2}}\left\{ {\lambda _1+\lambda _2{\rm sech}^2(\beta \tau
)}
\right.
\nonumber \\
+\left. {\lambda _3{\rm sech}^4(\beta \tau )+\lambda _4{\rm sech}^6
(\beta \tau )}
\right\}
\end{eqnarray}
where $\lambda _1=2v\left( {v^2+\beta ^2-1} \right)$,
$\lambda _2=2\alpha  +\alpha \beta ^2-3\beta ^2v-5\alpha v^2$,
$\lambda _3=4\alpha ^2v$ and
$\lambda _4=-\alpha ^3$.
We find that if $v$ satisfies
\begin{equation}
{{(v-\alpha)^3} \over v}-1 \ge 2\beta^2,
\end{equation}
then $U(\tau)$ in Eq. (10) is repulsive as required. Indeed,
$v$ has to be supersonic in order to satisfy the inequality (11).
Therefore we have provided an explicit solution of dressed
states at supersonic speeds.

In Fig. 1, we illustrate the general shapes of the matching
potentials $U(\tau)$ as a function of parameters $\alpha$ and
$\beta$. We plot a characteristic curve defined by the equality
sign in (11). Therefore the region under the curve corresponds to
purely repulsive potentials. These potentials have a maximum
barrier height at $\tau=0$. Above the curve, the second derivative
$U''(0)$ is positive which corresponds to an attractive force on
particles near the center of $U$. We shall not discuss the
solutions in such a domain, since the potentials are not purely
repulsive. We remark that at a higher speed $v$, the
characteristic curve takes a higher position on the $\alpha-\beta$
plane, allowing a wider range of parameters for the solution
(8-9). We also note that $U(\tau)$ takes a simpler form, $U(\tau)
\approx v \alpha {\rm sech}^2 (\beta \tau)$, when $v \gg \alpha,
\beta$.

There are interesting features in our solution: (I) A dressed
state for a {\em repulsive} $U$ at $v>1$ must come with a {\em
positive} $\alpha$. This means that the field concentrates at the
center of $U$, even though $U$ itself is repulsive. Such a
`counter-intuitive' behavior does not occur for dressed states in
the subsonic speed domain. It is worth noting that the family of
solutions Eqs. (8-10) can be generalized for subsonic speeds. In
that case, $\alpha$ is negative which recovers the expected dip at
the center of the potential. (II) The height of the matching
potential $U$ is bounded by
\begin{equation}
2U(0) \le (v^2-3v^{2/3} +2).
\end{equation}
This inequality sets an upper limit of $U(0)$ for the existence of
the solution (8-9). From the right side of Eq. (7), we see that a
large $U$ term would lead to an unbounded $r(\tau)$ unless the
$v^2$ term can make a balance. Hence for speeds $v$ which are
close to (but still greater than) $1$, the matching potential must
be weaker accordingly. It is also apparent that there are no
dressed states for impenetrable potentials.

In order to investigate the dynamical consequences of dressed
states, we solve numerically the time-dependent NLSE (1) for $\Psi
(x,t)$ which starts from the ground state $\Psi(x,0)=1$. The
potential $U(x-vt)$ is turned on suddenly at $t=0$. In Fig. 2a, we
plot the field energy, $\Delta E(t) = {1 \over 2}\int {dx\left(
{\left| {\nabla \Psi (x,t)} \right|^2 + \left| {\Psi (x,t)}
\right|^4 } \right)}  - E(0)$ relative to the initial field energy
$E(0)$ as a function of time. This quantity tells us how much
field energy is gained from the moving potential if it is switched
off at a time $t$. In Fig. 2a we see that there is a positive gain
of the field energy at early times. After a characteristic time of
order $2\beta^{-1}$, the energy eventually becomes a constant. In
this steady state regime, the moving potential does not `feel' any
the reaction force from the field even though $U$ is repulsive.

The emergence of a steady state indicates the formation of a
dressed state. This is explicitly shown in Fig. 2b where $|\Psi|$
are plotted as a function of $x$ at different times. The arrows in
the figure indicate the positions of the moving potential. We see
that the field develops a shape invariant wavepacket form at the
potential, i.e., the potential is dressed. By comparing both the
phase and the amplitude of the wavepacket with the analytic
expressions Eqs. (8) and (9), we found a very good agreement
between them. At regions far away from the potential, $\Psi$
exhibits features created by the sudden motion at $t=0$. We see
that the left-most wavefront propagates at the speed of sound, and
there is an oscillatory pattern developing in a depression region
near the middle. These features are partly responsible for the
transient energy gain shown in Fig. 2a. A closer look at the
depression region at longer times (not shown) indicates that the
short wavelength structure becomes grey solitons traveling at
subsonic speeds \cite{grey}.

We emphasize that the usual picture of critical velocities does
not apply to dressed states. The velocity-dependent requirements
here are specified by the inequalities (11) and (12), which do not
impose an upper limit for $v$. Instead, $v$ has to be sufficiently
high in order to achieve dressing for a given $U$ \cite{highv}. We
have calculated the energy gain using the {\em same} potential
shown in Fig. 2a but at {\em different} velocities. For velocities
near the speed of sound such that (12) is strongly violated, we
see no dressed state formation. In those cases the field energy
keeps growing via repeated emissions of grey solitons from the
potential as observed in Ref.\cite{hakim}.

Finally, we want to point out the stability of our solution. Since
the initial condition used in Fig. 2b is far away from the dressed
state solution, the convergence of the field precisely into the
analytical form (8) and (9) is an evidence of dynamic stability of
dressed states. We have performed direct stability tests based on
collective excitation frequencies in the co-moving frame. Our
numerical results confirm the dynamic stability of the solutions
\cite{remark}. The issue of the sensitivity to the shape of $U$
has also been examined. By replacing $U$ with a gaussian, we
repeat the numerical calculations as in Fig. 2b. As long as the
gaussian's width and height are well within the domain described
by inequalities (11) and (12), we see the emergence of dressed
states. Therefore dressed state formation is not sensitive to the
details of $U$.

To conclude, we have presented a family of motional dressed states
as exact analytic solutions of a nonlinear Schr\"odinger field
interacting with a moving potential. The formation of a dressed
state enforces frictionless motion, and so it preserves a key
feature in superfluid in the supersonic regime. We stress that our
method is not limited to the solution family (8-9) and potentials
in purely repulsive forms. With our approach, one can construct
wide varieties of solutions of the field and the corresponding
matching potentials, as long as the stability and physical
constraint for the potential are allowed. This opens interesting
possibilities of controlling local densities of a condensate
dynamically. Our study is for one-dimensional systems; we
speculate that similar dressing features appear in Bose
condensates in quasi-1D configurations, such as condensates
trapped in a long cigar shape or in a ring, within characteristic
time scales. It remains an important open question whether
dressing effects appear in two or higher dimensional systems. We
hope to address this issue in the future.

\acknowledgments We acknowledge the support from the Chinese
University of Hong Kong Direct Grant No. 2060148. C.K.L. is
supported by a postdoctoral fellowship at the Chinese University
of Hong Kong.


\begin{figure}
\centerline{
\epsfxsize=3.5in
\epsfbox{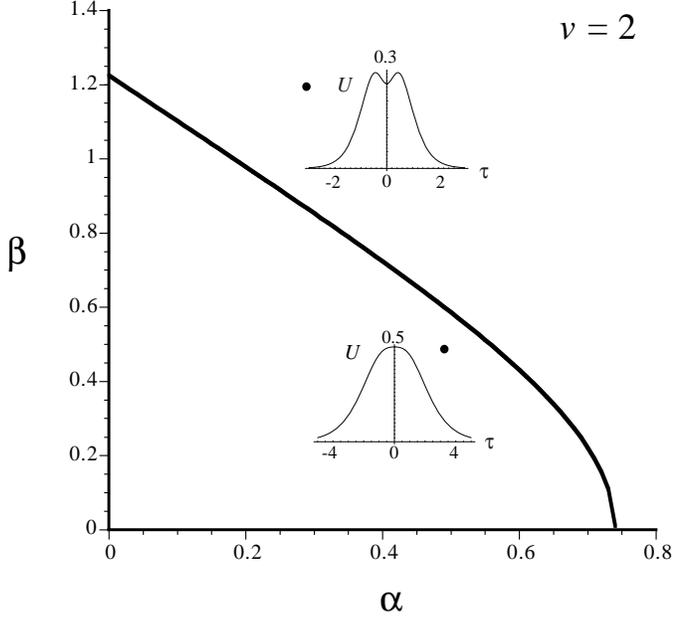}}
%\vspace{-10mm}
\caption{Parameter space of the solutions at a supersonic speed
$v=2$, with positive $\alpha$ and $\beta$. The characteristic
curve is defined by the equality sign of (11). The region under
the curve is associated with purely repulsive potentials $U$. The
shapes of $U$ at the two dots are representative examples above
and below the curve} \label{fig1}
\end{figure}

\begin{figure}
\centerline{
\epsfxsize=3.5in
\epsfbox{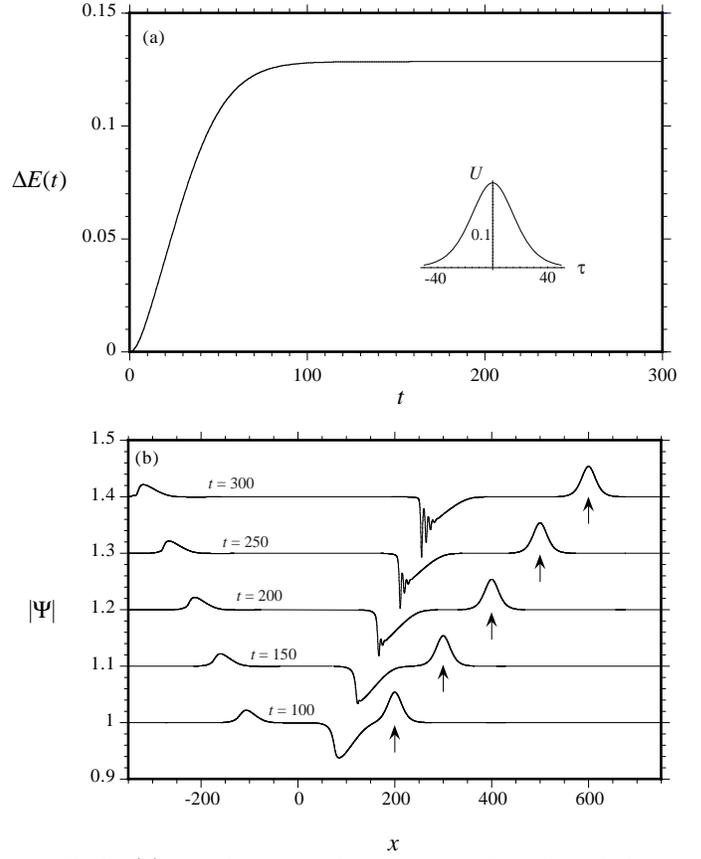}}
%\vspace{-10mm}
\caption{(a) The increase of energy as a function of time with the
parameters: $v=2$, $\alpha=0.2$ and $\beta=0.05$. The potential is
shown in the inset. (b) Time evolution of the modulus of $\Psi$ at
different times. The curves are shifted up by 0.1 successively in
order to display the evolution. The arrows indicate the position
of the center the potential at the corresponding time.}
\label{fig2}
\end{figure}


\begin{references}
\bibitem{highv} For example, in the case of $\alpha=1/v$,
$\beta=1$ and $v \gg 1$, the potential is approximately
independent of $v$, i.e., $U(\tau) \approx {\rm sech}^2 (\tau)$.
In this case, the dressed field solution $r(\tau) \approx 1+{\rm
sech}^2(\tau)/2v^2$ exists at any $v\gg 1$.
\bibitem{remark} In the co-moving frame $x'=x-vt$, $t'=t$,
the collective excitation frequencies with respect to a
stationary state $\psi$ (i.e., a dressed state) are
defined by the eigenvalues of the matrix,
\begin{eqnarray}
M = \left( {\matrix{
   {D + 2\left| \psi  \right|^2 } & {\psi ^2 }  \cr
   { - \psi ^{*2} } & { - D^* - 2\left| \psi  \right|^2 }  \cr
 } } \right)       \nonumber
\end{eqnarray}
where $D \equiv  - {1 \over 2}\partial_{x'}^2  + iv\partial
{_{x'}}$. The system is dynamically stable if all eigenvalues of
$M$ are real.
\end{references}
\end{document}